\documentstyle[sprocl]{article}
\def\Journal#1#2#3#4{{#1} {\bf #2}, #3 (#4)}
\def\NPB{{\em Nucl. Phys.} B}
\def\PLB{{\em Phys. Lett.}  B}
\def\PRL{\em Phys. Rev. Lett.}
\def\PRD{{\em Phys. Rev.} D}

\def\be{\begin{equation}}
\def\ee{\end{equation}}
\def\bea{\begin{eqnarray}}
\def\eea{\end{eqnarray}}

\begin{document}
\title{Effective Higgs-to-Light-Quark Coupling \\Induced by Heavy Quark
Loops}
\author{ M.R. AHMADY}
\address{Department of Physics, Ochanomizu University,\\
 1-1, Otsuka 2, Bunkyo-ku, Tokyo 112, Japan.}
\author{V. ELIAS, A.H. FARIBORZ, R.R. MENDEL \footnote{Deceased}}
\address{Department of Applied Mathematics, 
The University of Western Ontario, \\
 London, Ontario, Canada N6A 5B7.}
\maketitle\abstracts{
We study the coupling arising via perturbative QCD of zero 4-momentum
Higgs bosons to light quarks inside the nucleon.   
Qualitative comparison with the results obtained from 
one-loop-order low energy theorems for the Higgs-nucleon interaction
suggests the existence of a dynamical light-quark mass which falls
off with momentum.  Quantitative comparison leads to an estimate of
$\alpha_s$ at very low $q^2$ quite near unity.
}
\section{Introduction}
The coupling of the zero 4-momentum Higgs to nucleons is characterized by
a mass of order $210 MeV$.  Obviously, a direct coupling of the Higgs to 
a light quark inside nucleon is not sufficient to generate this
value.   The effective interaction of the Higgs with nucleons
is known to be enhanced via coupling of the nucleon to a heavy
quark triangle [Fig.1] through gluon exchanges.
A Higgs boson low energy theorem \cite{HNCoupling} has been applied in order
to estimate the effective coupling for this interaction. \cite{DawHab} This theorem
basically relates the matrix elements $ M(A \rightarrow B) $ and $M(A
\rightarrow B + Higgs) $ 
\begin{equation}
\lim_{p_{Higgs} \rightarrow 0} M(A \rightarrow B + Higgs )
=
{ 
   N_h \alpha_s ^2
   \over
   3\pi \langle \phi \rangle 
}
{
   \partial 
   \over
   \partial \alpha_s
}
M ( A \rightarrow B ), 
\label{eq:LET}
\end{equation}
where $N_h$ is the number of heavy quarks. If $A$ and $B$ are both
identified with the nucleon, then the physical matrix element $M (A
\rightarrow B )$ is the nucleon mass, which in the chiral limit is 
proportional to $\Lambda_{QCD}$.  Since  
$ \alpha_s =  (4 \pi )/\left[ 9 ln(p^2 / \Lambda_{QCD}^2) \right]$
to one loop order, we find that
\begin{equation}                                                    
\lim_{p_{Higgs} \rightarrow 0} M(q \rightarrow q + Higgs )
\equiv
g_{HNN}|_{induced}
=
  {{ 2 N_h m_N} \over { 27 \langle \phi \rangle }},
\label{eq:LET_Res}
\end{equation}
leading to an 
$N_h=3$ effective interaction characterized by a mass of about $210
MeV$. \cite{Okun}
This same argument can be applied to the (constituent-) quarks within
the nucleon.  In the chiral limit, the constituent-quark mass becomes
the dynamical mass associated with the chiral noninvariance of the
QCD vacuum.  The matrix element $M(A \rightarrow B)$ is just this
dynamical mass, $m_{dyn}$, which (like the nucleon mass $m_N$ in the
chiral limit) is
necessarily proportional to $\Lambda_{QCD}$ .
Consequently, one can repeat the derivation leading to
(\ref{eq:LET_Res}) for the
constituent-quark Yukawa interactions, and find the same result
except for the replacement of $m_N$ with $m_{dyn}$, corresponding to a
Yukawa-interaction mass of order $70 MeV$.
\begin{figure}[h] 
\vspace{1.2in}
\includegraphics{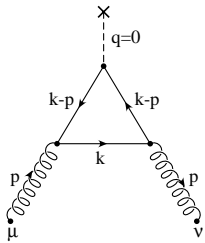}
\caption{The heavy quark triangle.}
\end{figure}

In this work we would like to estimate the Higgs coupling to the nucleon 
by performing an equivalent lowest order perturbative QCD 
calculation.  We study this coupling by coupling directly the heavy
quark triangle to the light quarks inside the nucleon [Fig.2]  and we
incorporate all non-perturbative QCD effects into a dynamical mass
function for the light quark.  Therefore in order to determine the
effective coupling we need to evaluate the two loop diagram in Fig.2. 
 As a first approximation,  we use a constant dynamical mass for the light
quark.  We subsequently utilize a more realistic  function of the
light quark momentum.

\begin{figure} 
\vspace{1.5in}
\includegraphics{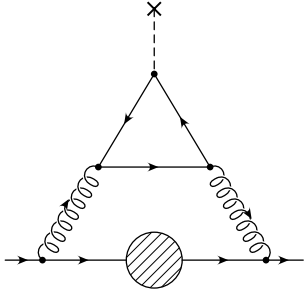}
\caption{The two loop diagram representing the coupling of a zero 
4-momentum Higgs to a light quark through
gluons via a heavy quark triangle.}
\end{figure}

The heavy quark triangle \cite{triangle} [Fig.1] for heavy-quark mass $M$ and 
gluon-momentum $p$ couples the zero 4-momentum Higgs to two gluons, 
and is given by the following tensor
\begin{equation}
I_{\mu \nu }^{ab}= 
{
{2 \alpha}
\over 
{\pi \langle \phi \rangle }
}
\delta ^{ab} (p_\mu p_\nu -p^2 g_{\mu \nu }) 
\left[ 
{M^2 \over p^2} +       {
                            {2M^4} 
                            \over 
                            {p^4 \sqrt { 1- 
                                             {
                                                 4 M^2 
                                                 \over
                                                 p^2 
                                             }
                                       }
                            }
                        }
                          ln | {{\tau _+} \over {\tau _-}}|
\right],
\label{eq:Triangle} 
\end{equation}
where $ \tau _{ \pm } \equiv 1 \pm \sqrt { 1- 4 M^2 /p^2}$, $\alpha_s
=g_s^2/(4\pi)$, and $\langle \phi \rangle $ is the
vacuum expectation value of the Higgs field.  
$I_{\mu \nu }^{ab}$ is completely transverse which ensures gauge-parameter
independence of light-quark Yukawa couplings induced via Fig.2.
The expression in square brackets in (\ref{eq:Triangle}) goes to $-1/6$ in the 
heavy-quark limit ($M^2 \gg p^2$). For very large gluon momenta
($p^2$) the expression in brackets
vanishes, demonstrating that the heavy quark triangle serves as a 
cut-off in the integrals over the gluon momenta of Fig.2.  This
property ensures that the triangle of Fig.1 will not lead to
divergent renormalization-dependent results when incorporated into
Fig.2.
\section{The Effective Coupling to a Light Quark with Constant Dynamical Mass}
The Higgs coupling  to light quarks induced via Fig.2 may be 
expressed as 
\begin{equation}
\Sigma _{ind}(k) = 
\Sigma _0 (k^2) + k \! \! \! / \Sigma _1 (k^2).
\label{eq:Sigma_ind}
\end{equation}
To estimate $\Sigma_0(k^2)$ and $\Sigma_1(k^2)$, we first evaluate the
two loop diagram in Fig.2 with a constant dynamical mass.   We find 
it convenient to express the heavy quark triangle 
(\ref{eq:Triangle}) in the
following form:
\begin{equation}
I_{\mu \nu }^{ab}(p)= 
{
   {4 \alpha _s M^2 }
   \over
   {\pi \langle \phi \rangle }
}
\delta ^{ab} (p_\mu p_\nu -p^2 g_{\mu \nu }) 
\int _0^1 dy 
{
   y
   \over
   { p^2 - 
          {
            M^2
            \over
            {y(1-y)}
          }           
    }
}. 
\label{eq:Triangle_2}
\end{equation}
We find that
\begin{equation}
\Sigma _0 (k^2) = 
{
   {4 \alpha_s ^2 m_{dyn} }
\over 
{\pi ^2 \langle \phi \rangle }
} 
\int _0^1 y dy \int _0^1 dx_1 \int_0^{1-x_1} dx_2 L_0^{-1},
\label{eq:H_S0}                                                 
\end{equation}
\[
\Sigma _1 (k^2) = 
{
   {4 \alpha_s ^2  }
\over 
{\pi ^2 \langle \phi \rangle }
} 
\int _0^1 y dy \int _0^1 dx_1 \int_0^{1-x_1} dx_2 (1-x_1-x_2)
\]
\begin{equation}
\times 
\left[
(3x_2-1) L_0^{-1} + k^2 x_2^2 (1-x_2) M^{-2} L_0^{-2}
\right],
\label{eq:H_S1}                                          
\end{equation}
where
\begin{equation}
L_0=x_1 \left[ y (1-y) \right] ^{-1} +k^2 M^{-2} x_2 (x_2-1) +m^2
M^{-2} x_2.
\end{equation}
In the limit where $m_{dyn} \ll M$ 
\begin{equation}
\Sigma_0 (0) \rightarrow 
   2 \alpha _s^2 m_{dyn} \left[ ln (M/m_{dyn})+5/6 \right]/
(3\pi ^2 \langle \phi \rangle ),
\label{eq:H_S0_0}
\end{equation}
\begin{equation}
\Sigma_1 (0) \rightarrow 
  - \alpha _s^2/(18 \pi ^2 \langle \phi \rangle ),
\label{eq:H_S1_0}
\end{equation}
which indicates that $\Sigma_0(0)$ in the limit of very heavy masses
in the triangle depends logarithmically on the
heavy quark mass.  On the other hand $\Sigma_1(0)$ is independent of
the heavy quark mass in this limit.   These properties also apply
when the light quark has nonzero momentum $k$ [Fig.'s 3 and 4].  
Heavy flavours have different contributions in $\Sigma_0$, but have
almost the same contributions in $\Sigma_1$.
For a dynamical mass of $300 MeV$, we have  
$\Sigma_0 (0)\approx 268  \alpha_s^2 MeV/\langle \phi \rangle$ and
$\Sigma_1 (0)\approx -0.05 \alpha_s^2 /\langle \phi \rangle$, and if we
identify the $ k \! \! \! /$ term with the dynamical mass, we
have $\Sigma_{ind}(0) \approx 253 \alpha_s^2 MeV/\langle \phi
\rangle$.  This result coressponds to an induced Yukawa-coupling mass
of $70 MeV$, consistent with the one-loop low-energy theorem result
(\ref{eq:LET_Res}), provided $\alpha_s \approx 0.53$.   However, the
result clearly is dependent through (\ref{eq:H_S0_0}) upon the heavy
quark mass $M$, inconsistent with the low-energy-theorem result
(\ref{eq:LET_Res}).
\section{The Effective Coupling to a Light Quark with a Dynamical Mass
Function} 
We now consider a more realistic case in which the light quark has 
a momentum-dependent dynamical mass, and is therefore more sensitive to
infrared dynamics corresponding to the low energy region for
$\alpha_s$.  We use the dynamical mass function proposed by 
Holdom \cite{Holdom}
\begin{equation}
\Sigma_{QCD}(p^2)=
   (A+1)\Lambda^3 /( A\Lambda^2-p^2 ),
\label{eq:Sigma_QCD}
\end{equation}
where $A$ is a constant and $\Lambda$ has the dimensions of energy and 
is the low energy mass scale.  This expression has been
successfully tested with different physical parameters at low
energies. \cite{Holdom}  We find that 
\[
\Sigma_0(k^2) = 
{
{4 \alpha_s^2 M^2}
\over 
{3i\pi^4\langle \phi \rangle }
}
\int_0^1 y dy \int d^4p 
\]
\begin{figure}
\vspace{2.5in}
\includegraphics{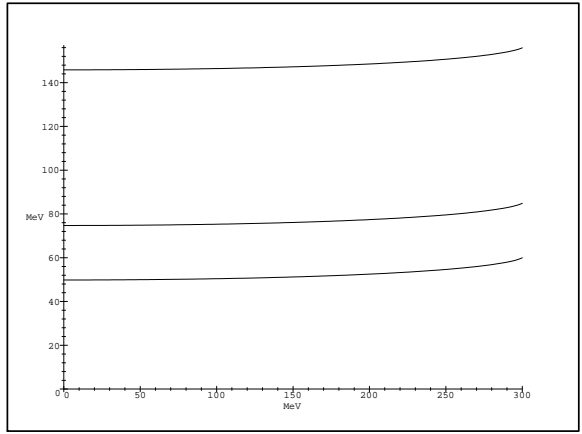}
\caption{Contributions of top, bottom, and charm quarks to $\Sigma_0 (k^2)$ 
in units of $ \alpha_s^2 /\langle \phi \rangle $ versus $(k^2)^{1/2}$
with a  constant dynamical mass of $300 MeV$ for the light quark. 
The top quark has the largest contribution,
followed by bottom and charm quark contributions respectively.}
\end{figure}
\begin{figure}
\vspace{2.5in}
\includegraphics{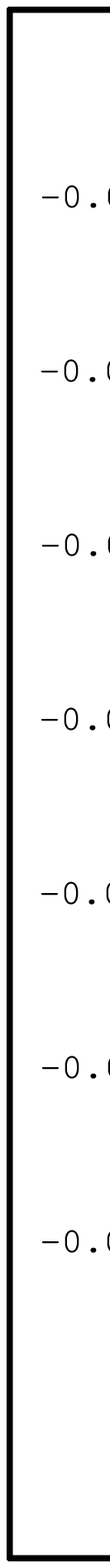}
\caption{Contributions of top, bottom, and charm quarks to 
$\Sigma_1 (k^2)$ 
in units of $ \alpha_s^2/\langle \phi \rangle $ versus $(k^2)^{1/2}$
with a constant dynamical mass of $300 MeV$ for the light quark.  
The top quark has the largest contribution,
followed by bottom and charm quark contributions respectively.}
\end{figure}
\begin{equation}
\times { 
{\gamma_\mu  (A+1)\Lambda^3  \left[ A\Lambda^2-(p-k)^2 \right]
\gamma_\nu (p^\mu p^\nu -p^2g^{\mu \nu})  
} 
\over
{\left[ (p-k)^2 \left[ A\Lambda^2 -(p-k)^2 \right] ^2 -(A+1)^2\Lambda
^6 \right] p^4 \left( p^2 - {M^2 \over {y(1-y)}} \right)}                                                              
},
\label{eq:DM_S0}
\end{equation}
\[
k \! \! \! / \Sigma_1(k^2)  = 
{
{4 \alpha_s ^2 M^2}
\over 
{3i\pi^4 \langle \phi \rangle }
}
\int _0^1 ydy \int d^4p 
\]
\begin{equation}
\times {
   {\gamma_\mu (p\! \! \! / -k\! \! \! /)
   \left[A \Lambda^2 -(p-k)^2\right]^2\gamma_\nu (p^\mu p^\nu-p^2g^{\mu
   \nu } )}
   \over 
   {
    \left[ 
          (p-k)^2 \left[ A\Lambda^2 -(p-k)^2 \right]^2
          -(A+1)^2\Lambda^6
    \right] 
            p^4 
                 \left( p^2-  {
                                M^2 
                                \over 
                                y(1-y)
                               }
                 \right) 
   } 
}.              
\label{eq:DM_S1}                                                 
\end{equation}
For a nonzero light-quark momentum $k$, these loop integrals are 
quite complicated. However, for $k^2 = 0$,  the quark's Lagrangian-mass 
shell in the chiral limit,  we find that
\[
\Sigma_0(0) = 
{
{4 \alpha ^2  M^2 (A+1) \Lambda ^3 } 
\over
{\pi^2 \langle \phi \rangle}
}
\int_0^1 y dy \int _0^{\infty } dx
(A \Lambda ^2 + x )
\]
\begin{equation}
\times
\left[ \left[ x ( A \Lambda ^2 +x )^2 + ( A + 1 )^2 \Lambda ^6 \right] 
\left( x + M^2/[y(1-y)] \right) \right]^{-1},
\label{eq:DM_S0_0}
\end{equation}
\[
\Sigma_1(0) = 
{
{2 \alpha ^2  M^2 (A+1)^2 \Lambda ^6 } 
\over
{\pi^2 \langle \phi \rangle}
}
\int_0^1 y dy \int _0^{\infty } dx
(3 x^2 + 4 A \Lambda ^2 x + A^2 \Lambda ^4 )
\]
\begin{equation}
\times
\left[ 
       \left[ x ( A \Lambda ^2 +x )^2 + ( A + 1 )^2 \Lambda ^6 \right] ^2
                         \left( x + M^2 /[y(1-y)] \right) 
\right] ^{-1}.
\label{eq:DM_S1_0}
\end{equation}                                                         
Choosing $A = 2$ and $\Lambda = 317 MeV$, \cite{Holdom} we find that 
$ \Sigma_0(0) = 62 \alpha_s ^2 MeV / {\langle \phi \rangle}$
and $\Sigma_1(0)=0.051 \alpha_s^2/\langle \phi \rangle $,
in which case $\Sigma_{ind}\approx (62 \rightarrow 77) MeV$ 
depending upon whether we consider $k \! \! \! /$ on the dynamical $O
(300 MeV)$ mass shell \cite{HillElias}
or assign $k \! \! \! /$ a value of zero
(consistent with the vanishing Lagrangian
mass in the chiral limit).  In either case, however, we obtain a
result that is independent of the heavy quark mass, as predicted by
the low energy therorem (\ref{eq:LET}). Quantitative agreement with
(\ref{eq:LET_Res}) is obtained  
provided $\alpha_s \approx 0.95 \rightarrow 1.06$.
The near unity value of $\alpha_s$ is consistent with 
infrared expectations, particularly criticality arguments \cite{Higashijima,ChiralSB}
for chiral 
symmetry breaking  as well as the anticipated freezout of the 
strong coupling near unity discussed by
Mattingly and Stevenson. \cite{LEStudy}

The momentum dependence of the effective Yukawa interaction can be
studied via the following approximation in the denominator of the momentum 
integrals (\ref{eq:DM_S0}) and (\ref{eq:DM_S1})
\begin{equation}
\Sigma_{QCD}(k-p)^2 \rightarrow \Sigma_{QCD}(0),
\label{eq:Sigma_QCD_lim}
\end{equation}
an approximation providing a lower bound for the effective Yukawa
interaction.  Making this substitution, we find that
\begin{equation}
\Sigma_0 (k^2) = {{-4(A+1) \alpha ^2 M^2}\over {\Lambda \pi ^2 <\phi
>}}\int _0^1 y dy \int _0^1 dx_1 \int _0^{1-x_1}dx_2 \int
_0^{1-x_1-x_2}dx_3  L^{-2},
\label{eq:DM_S0_lim}
\end{equation}
\[
\Sigma_1 (k^2) = {{8 \alpha ^2 M^2}\over {\Lambda
^2 \pi ^2 <\phi
>}}\int _0^1 y dy \int _0^1 dx_1 \int _0^{1-x_1}dx_2 \int
_0^{1-x_1-x_2}dx_3 
\]
\begin{equation}
\times
(1-x_1-x_2-x_3)
\left[ 
       (1/3) F_0 \Lambda ^{-4} L^{-3}
       - (1/3) F_1 \Lambda ^{-2} L^{-2} 
       + F_2 L^{-1}
\right],
\label{eq:DM_S1_lim}
\end{equation}
where 
\begin{equation}
L={{M^2x_3}\over {\Lambda ^2 y(1-y)}}-{k^2\over \Lambda
^2}(x_1+x_2)(1-x_1-x_2)+{\left[ {A+1}\over A\right] ^2}x_1+Ax_2,
\label{eq:L}
\end{equation}
\[
F_0=3 k^2 (x_1+x_2)^2 \left[ k^2 - A \Lambda^2 + 3 k^2 (x_1+x_2)
(x_1+x_2 -1) + \right.
\]
\begin{equation}
\left. A \Lambda ^2 (x_1+x_2) - k^2 (x_1+x_2)^3 \right],
\label{eq:F_0}
\end{equation}
\[
F_1=-{3\over 2} (A \Lambda^2 -k^2) + (x_1+x_2) \left[ (9/2)( A
\Lambda^2 -k^2 )- 12 (x_1+x_2)^2 k^2 +\right.
\]
\begin{equation}
\left. 21 (x_1+x_2)k^2 -6 k^2 \right],
\label{eq:F_1}
\end{equation}
\begin{equation}
F_2=-6 (x_1+x_2) + 3.
\label{eq:F_2}
\end{equation}
In this ``lower-bound'' approximation, the momentum dependence of $\Sigma_0 (k^2)$ and $\Sigma_1 (k^2)$
[Fig.'s 5 and 6] confirm the heavy flavour independence of the
induced coupling in the heavy quark limit (the contributions of top and
bottom quarks lead to almost identical curves).  
As evident from these figures, there is
a large enhancement in the coupling around $k^2 \approx (500 MeV)^2$.
This is due to the onset of branch cut singularity beginning when $k^2
= \left[ \Sigma_{QCD}(k^2) \right]^2 $ which occurs when
$\sqrt{k^2} \approx 530 MeV$.  However, caution must be used in
attempting a physical interpretation for this branch cut, because
the form chosen for $\Sigma_{QCD}$ is not likely to be applicable for 
Minkowskian $k^2$ near or past the singularity at $k^2=A \Lambda ^2$.  

The ``lower-bound'' approximation indicates only a soft dependence
of $\Sigma_0$ on $k^2$ near the origin [Fig. 5].  This can be
tested by considering the Taylor series for $\Sigma_0(k^2)$
explicitly:
\[
\Sigma_0(k^2)=\Sigma_0(0)+ 
{1\over 2}
\left[ 
{\partial ^2 \over 
{ \partial k^\rho \partial k^\sigma }}
\Sigma_0 (k^2) 
\right]_{k=0}
k^{\rho} k^{\sigma }
+ O (k^2)^2
\]
\[
\approx
\Sigma_0(0) +
{{4 \alpha ^2 M^2 (A+1) \Lambda ^3 }\over { \pi ^2 \langle \phi
\rangle }} k^2
\int _0^1 y dy \int _0^\infty dx 
\left(
-x^6-3 A \Lambda^2 x^5 - 3 A^2 \Lambda ^4 x^4 + 
\right.
\]
\[
\Lambda ^6 x^3 \left[ 7(1+A)^2 -A^3 \right] +12 A (1+A)^2 \Lambda^8 x^2
+ 6 A^2 (1+A)^2 \Lambda^{10} x + (1+A)^2 
\]
\begin{equation}
\times \left.\left[ A^3-(1+A)^2\right] \Lambda^{12}
\right)
\left[
       \left( x +   M^2 /[ y ( 1 - y )] 
       \right)
\left[
x ( A\Lambda^2+x)^2 +
\Lambda^6(1+A)^2 
\right]^3
\right]^{-1}
\label{eq:DM_S0_Taylor}
\end{equation}
\begin{figure}
\vspace{2.5in}
\includegraphics{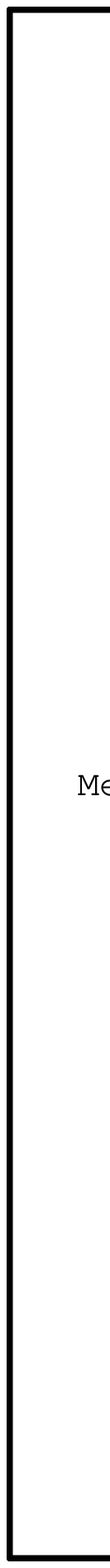}
\caption{Lower-bound-approximation contributions of top, bottom, and charm quarks to $\Sigma_0 (k^2)$ 
in units of $ \alpha_s^2/\langle \phi \rangle $ versus $(k^2)^{1/2}$ 
with the dynamical mass function $\Sigma_{QCD}$ for the light quark.
The top quark has the largest contribution, followed by bottom and charm 
quark contributions respectively.}
\end{figure}
\begin{figure}
\vspace{2.5in}
\includegraphics{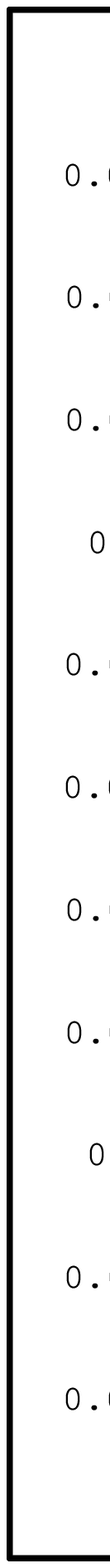}
\caption{Lower-bound-approximation contributions of top, bottom, and charm quarks to $\Sigma_1 (k^2)$ 
in units of $ \alpha_s^2/\langle \phi \rangle $ versus $(k^2)^{1/2}$ 
with the dynamical mass function $\Sigma_{QCD}$ for the light quark.
The top quark has the largest contribution, followed by bottom and charm 
quark contributions respectively.}
\end{figure}
\begin{figure}
\vspace{2.5in}
\includegraphics{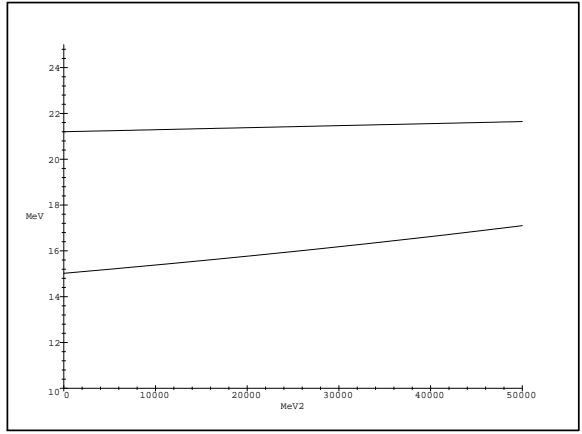}
\caption{Contribution of the top quark to the linear approximation and 
lower-bound approximation for $\Sigma_0 (k^2)$ in units of $\alpha_s^2/\langle \phi
\rangle $ versus $k^2$, assuming the dynamical mass function $\Sigma_{QCD}$
for the light quark.}
\end{figure}
The next-to-leading correction in (\ref{eq:DM_S0_Taylor}) leads to a linear approximation
to $\Sigma_0(k^2)$ whose slope is consistent with the lower bound
approximation as $k^2 \rightarrow 0$, as shown in Fig.7. Finally, we
note that the induced coupling on the dynamical {\it Euclidean} mass shell of
the light quark at $k^2=-\Lambda^2$ is found from
(\ref{eq:DM_S0_lim}) and (\ref{eq:DM_S1_lim}) to be $\Sigma_{ind}(-\Lambda
^2) \ge 48 \alpha_s^2 MeV/\langle \phi \rangle $,
a result in agreement with the low energy theorem (\ref{eq:LET}) 
provided $\alpha_s \le 1.2$.

\section{Discussion}
By using a realistic dynamical mass function (\ref{eq:Sigma_QCD}),
we obtain a heavy-quark-induced Yukawa interaction with nucleonic
constituent quarks that is independent of the heavy quark mass, as predicted by
the Higgs low-energy theorem. Quantitative agreement with the 
low-energy theorem result is obtained provided the infrared value of 
$\alpha_s $ is $0.95 \rightarrow 1.06$, a range
anticipated from criticality arguments for chiral symmetry breaking
and from the expected freezout of the strong coupling near unity.

We reiterate that the low-energy theorem (\ref{eq:LET_Res}) follows from 1-loop
corrections to the gluon propagator, motivating our explicit
comparison to the lowest order graph [Fig.2] in perturbative QCD. 
The large size of $\alpha_s$ that follows from such a comparison
suggests a need to consider three-loop diagrams as well.  However,
there are reasons to believe that the relevant expansion parameter
near criticality is $\alpha_s N_c /(4 \pi)$, perhaps providing some suppression of
higher-order contributions. \cite{Near_Criticality}
 
\section*{Acknowledgments}
We are grateful for extensive discussions with V. A. Miransky, and to
the Natural Science and Engineering Research Council of Canada for
financial support.  We also note with sorrow that R. R. Mendel, who
initiated this research, was actively pursuing its resolution at the
time of his tragic death last August. 

\end{document}